\documentstyle[12pt,psfig]{article}
\newcommand{\mysection}{\setcounter{equation}{0}\section}

\def\beq{\begin{equation}}
\def\eeq{\end{equation}}
\def\beqa{\begin{eqnarray}}
\def\eeqa{\end{eqnarray}}

\begin{document}

\begin {flushright}
FSU-HEP-991123\\
CERN-TH/99-339\\
DFTT 62/99\\
\end {flushright} 
\vspace{3mm}
\begin{center}
{\Large \bf Electroweak-boson hadroproduction at large transverse momentum: 
factorization, resummation, and NNLO corrections}
\end{center}
\vspace{2mm}
\begin{center}
{\large Nikolaos Kidonakis}\\
\vspace{2mm}
{\it Department of Physics\\
Florida State University\\
Tallahassee, FL 32306-4350, USA} \\
\vspace{8mm}
{\large Vittorio Del Duca}\footnote{On leave from INFN, Sezione di Torino,
Italy}\\
{\it Theory Division, CERN\\ CH-1211
Geneve 23, Switzerland}
\vspace{2mm}
\end{center}

\begin{abstract}
We study the resummation of singular distributions 
for electroweak-boson production at large
transverse momentum in hadronic collisions. 
We describe the factorization properties of the cross section near 
partonic threshold and construct the resummed cross section 
at next-to-leading logarithmic accuracy
in moment space in terms of soft gluon anomalous dimensions.
We present full analytical results for the expansion of the 
resummed cross section up to next-to-next-to-leading order.
Our results can be applied to $W$, $Z$, and virtual $\gamma$
production at hadron colliders.
 
\end{abstract}
\vspace{2cm}

\begin{flushleft} CERN-TH/99-339 \\
November 1999
\end{flushleft}

\pagebreak

\mysection{Introduction}

The study of electroweak-boson production in hadron colliders  
provides useful tests of the standard model and estimates of
backgrounds to new physics. For example, $W b {\bar b}$ production
is the principal background to the associated Higgs boson
production, $p{\bar p}\to H(\to b {\bar b}) W$, at the Tevatron \cite{EV}. 
The complete calculation of the next-to-leading-order (NLO)
cross section for $W$, $Z$, and virtual $\gamma$ production
at large transverse momentum in hadron collisions has been
presented in Refs. \cite{AR,gpw}, following earlier
results for the non-singlet cross section in Ref. \cite{emp}.   

The calculation of hard scattering cross sections near the edges
of phase space (partonic threshold), such as electroweak-boson
production at high transverse momentum, involves corrections from the 
emission of soft gluons from the partons in the process. 
At each order in perturbation theory one encounters large 
logarithms that arise from incomplete cancellations near partonic
threshold between graphs with real emission and virtual graphs.
These threshold corrections exponentiate and have been resummed 
explicitly at next-to-leading logarithmic accuracy for a number 
of processes including heavy quark~\cite{KS,NKJSRV,NKRV,bcmn,corfu,LM}, 
dijet~\cite{KOS,corfu}, and direct photon production~\cite{LOS,cmn,NKJO}, 
using general techniques developed originally for Drell-Yan 
production \cite{ster,CT1}. For a review see Ref. \cite{NK}.

In this paper we study electroweak-boson production at large
transverse momentum in hadronic collisions where these considerations
are of relevance. Following Ref. \cite{LOS}, we discuss in Section 2
the factorization properties of the single-particle inclusive
cross section \cite{conto,KS,KOS,LOS} and identify its singular 
behavior near threshold. 
We then proceed to resum the leading (LL) and next-to-leading (NLL) 
logarithms explicitly to all orders in perturbation theory.
In Section 3 we provide full analytical
results for the expansion of the resummed cross section at 
NLO and at next-to-next-to-leading order (NNLO).
Our NLO expansion agrees near partonic threshold with the 
exact NLO calculations in Refs. \cite{AR,gpw} while our NNLO results provide 
new predictions.
 
\mysection{Factorization and resummation}

For the hadronic production of an electroweak boson $V$ of mass $m_V$,
where $V=W,Z,\gamma^*$,
\beq
h_A(P_A)+h_B(P_B) \longrightarrow V(Q) + X \, ,
\eeq
we write the factorized single-particle cross section as
\beqa
E_Q\,\frac{d\sigma_{h_Ah_B{\rightarrow}V(Q)+X}}{d^3Q} &=&
\sum_{f} \int dx_1 \, dx_2 \; \phi_{f_a/h_A}(x_1,\mu_F^2) 
\; \phi_{f_b/h_B}(x_2,\mu_F^2) 
\nonumber \\ && \hspace{-10mm} \times \,
E_Q\,\frac{d\hat{\sigma}_{f_af_b{\rightarrow}V(Q)+X}}{d^3Q}
(s,t,u,Q,\mu_F,\alpha_s(\mu_R^2)) \label{factor}
\label{factV}
\eeqa
where $E_Q=Q^0$, $\phi_{f/h}$ is the parton distribution for parton 
$f$ in hadron $h$,
and $\hat{\sigma}$ is the perturbative cross section. The initial-state 
collinear singularities are factored into the
$\phi$'s at factorization scale $\mu_F$, while $\mu_R$ is the
renormalization scale.

At the parton level, the subprocesses for the 
production of an electroweak boson and a parton are 
\beqa
q(p_a)+g(p_b) &\longrightarrow& q(p_c) + V(Q) \, ,
\nonumber \\
q(p_a)+{\bar q}(p_b) &\longrightarrow& g(p_c) + V(Q) \, .
\label{partsub}
\eeqa
The hadronic and partonic kinematic invariants in the process are 
\beqa
&& \hspace{-5mm} S=(P_A+P_B)^2, \; T=(P_A-Q)^2, \; U=(P_B-Q)^2, \;
S_2 = S + T + U - Q^2,
\nonumber \\ 
&& \hspace{-5mm} s=(p_a+p_b)^2, \; t=(p_a-Q)^2, \; u=(p_b-Q)^2, \;
s_2 = s + t + u - Q^2,
\label{partkin} 
\eeqa
where $S_2$ and $s_2$ are the invariant masses of the system recoiling 
against the electroweak boson at the hadron and parton levels, respectively.
$s_2=(p_a+p_b-Q)^2$ parametrizes the inelasticity of the parton scattering: 
for one-parton production, $s_2=0$.
Since $x_i$ is the initial-parton momentum fraction, defined
by $p_a = x_1 P_A$ and $p_b = x_2 P_B$, hadronic and partonic kinematic 
invariants are related by 
$s = x_1 x_2 S$, $t-Q^2 = x_1 (T-Q^2)$, and $u-Q^2 = x_2 (U-Q^2)$.

In general, $\hat{\sigma}$ includes distributions with respect 
to $s_2$ at $n$th order in $\alpha_s$ of the type
\beq
\left[{\ln^{m}(s_2/Q^2) \over s_2} \right]_+, \hspace{10mm} m\le 2n-1\, ,
\label{zplus}
\eeq
defined by their integral with any smooth function $f$ by 
\beq
\int_0^{Q^2} ds_2 \, f(s_2) \left[{\ln^m{(s_2/Q^2)}\over s_2}\right]_{+} =
\int_0^{Q^2} ds_2 {\ln^m{(s_2/Q^2)}\over s_2} [f(s_2) - f(0)]\, .
\label{splus}
\eeq
These plus distributions are the remnants of cancellations between
soft and virtual contributions to the cross section. 

We now consider the partonic cross section, with the colliding hadrons in 
Eq. (\ref{factor}) replaced by partons. To organize the plus distributions
in ${\hat \sigma}$, we introduce a refactorization \cite{KS,KOS,LOS}
using new functions $\psi_{f/f}$ and $J$ that describe the dynamics of partons 
moving collinearly to the incoming and outgoing partons respectively, and
functions $H$ and $S$ which describe respectively the dynamics of hard 
partons and of soft partons which are not collinear to $\psi_{f/f}$ and $J$.
This factorization is shown in Fig. 1 in cut diagram notation,
showing contributions from the amplitude and its complex conjugate,
with $H=hh^*$. 
At partonic threshold we find the relation
\beqa
\frac{S_2}{S} &\simeq& -(1-x_1) {u\over s} - (1-x_2) {t\over s} 
+ {s_2\over s} \nonumber\\
&\equiv& w_1 \left(\frac{-u}{s}\right) +w_2\left(\frac{-t}{s}\right)  
+ w_J + w_s\, ,
\label{sdue}
\eeqa
where in the second line of the equation the weights $w$ identify
the contributions of the functions in the refactorized cross section.
At fixed $S_2$, the partonic cross section factorizes into
\beqa
&& \hspace{-10mm} E_Q\,\frac{d{\sigma}_{f_af_b\rightarrow V}}{d^3Q}
= H \int dw_1 dw_2 dw_J dw_s\, \psi_{f_a/f_a}(w_1)\, 
\psi_{f_b/f_b}(w_2)\, J(w_J)\, S(w_sQ/\mu_F) 
\nonumber \\ && \times \,
\delta\left({S_2\over S} - w_1\left(\frac{-u}{s}\right)   
- w_2\left(\frac{-t}{s}\right) - w_J - w_s \right)\, .
\label{softfact}
\eeqa
If we take moments of the above equation, with $N$ the moment
variable, we can write the partonic cross section as
\beqa
\int \frac{dS_2}{S} {\rm e}^{-NS_2/S}
E_Q\frac{d{\sigma}_{f_a f_b\rightarrow V}} {d^3 Q}
&=&H \,\int dw_1 {\rm e}^{-N_1 w_1} {\psi}_{f_a/f_a}(w_1)\, 
\nonumber \\ && \hspace{-50mm} \times \,
\int dw_2 {\rm e}^{-N_2 w_2} {\psi}_{f_b/f_b}(w_2)\,
\int dw_J {\rm e}^{-N w_J} J(w_J)\,
\int dw_s {\rm e}^{-N w_s} S(w_sQ/\mu_F)
\nonumber \\ && \hspace{-30mm}
\equiv{\tilde{\psi}}_{f_a/f_a}(N_1)\, {\tilde{\psi}}_{f_b/f_b}(N_2)\,
{\tilde{J}}(N) \, H \, {\tilde{S}}(Q/(N\mu_F)) \, ,
\label{sigmamomdp}
\eeqa
with $N_1=N(-u/s)$ and $N_2=N(-t/s)$ 
for either partonic subprocess.

Taking moments of Eq. (\ref{factV}) with the incoming hadrons
replaced by partons, and using the first line of Eq. (\ref{sdue}),
we also have the relation
\beqa
\hspace{-1cm}
\int \frac{dS_2}{S} {\rm e}^{-NS_2/S}
E_Q\frac{d{\sigma}_{f_af_b\rightarrow V}}{d^3 Q}&=& 
\int dx_1 \, {\rm e}^{-N_1(1-x_1)} \phi_{f_a/f_a}(x_1,\mu_F^2) \,
\nonumber \\ && \hspace{-55mm} \times
\int dx_2 \, {\rm e}^{-N_2(1-x_2)} \phi_{f_b/f_b}(x_2,\mu_F^2) \,
\int \frac{ds_2}{s} \, {\rm e}^{-Ns_2/s} E_Q
\frac{d{\hat \sigma_{f_af_b\rightarrow V}}(s_2)}{d^3 Q}
\nonumber \\ 
&\equiv& {\tilde\phi}_{f_a/f_a}(N_1) {\tilde\phi}_{f_b/f_b}(N_2) E_Q 
\frac{d{\hat \sigma_{f_af_b\rightarrow V}}(N)}{d^3 Q} \, .
\label{moms4}
\eeqa
Note that $s_2/s=s_2/S$ up to quadratic terms in $(1-x_1)$ and/or $(1-x_2)$ 
and/or $s_2$.
Comparing Eqs. (\ref{moms4}) and (\ref{sigmamomdp}) we then may
solve for the moments of the perturbative cross section ${\hat\sigma}$:
\beq
E_Q\frac{d{\hat \sigma}_{f_a f_b \rightarrow V}(N)}{d^3 Q}
=\frac{{\tilde{\psi}}_{f_a/f_a}(N_1)\, {\tilde{\psi}}_{f_b/f_b}(N_2)}
{{\tilde{\phi}}_{f_a/f_a}(N_1)\,  {\tilde{\phi}}_{f_b/f_b}(N_2)}
{\tilde{J}}(N) \, H \, {\tilde{S}}(Q/(N\mu_F)) \, .
\label{nresum}
\eeq
The plus distributions in 
$\hat \sigma$ produce, under moments, powers of $\ln N$ as high
as $\ln^{2n} N$.
The LL corrections are included in  $\psi/\phi$ and $J$,
while the NLL corrections are included in $\psi/\phi$, $J$, and $S$.

\begin{figure}
\centerline{
\psfig{file=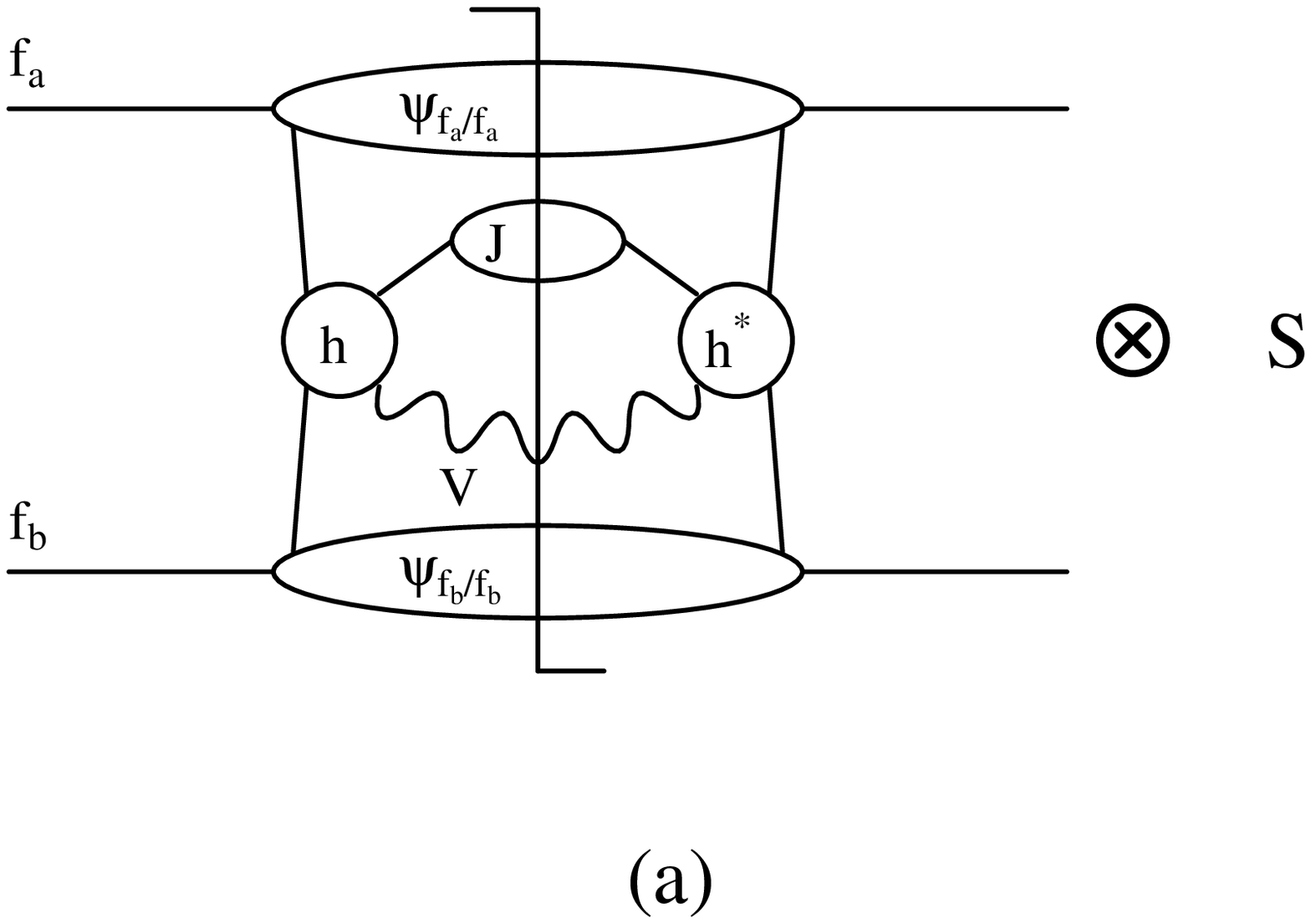,height=1.7in,width=2.5in,clip=}
\hspace{8mm}
\psfig{file=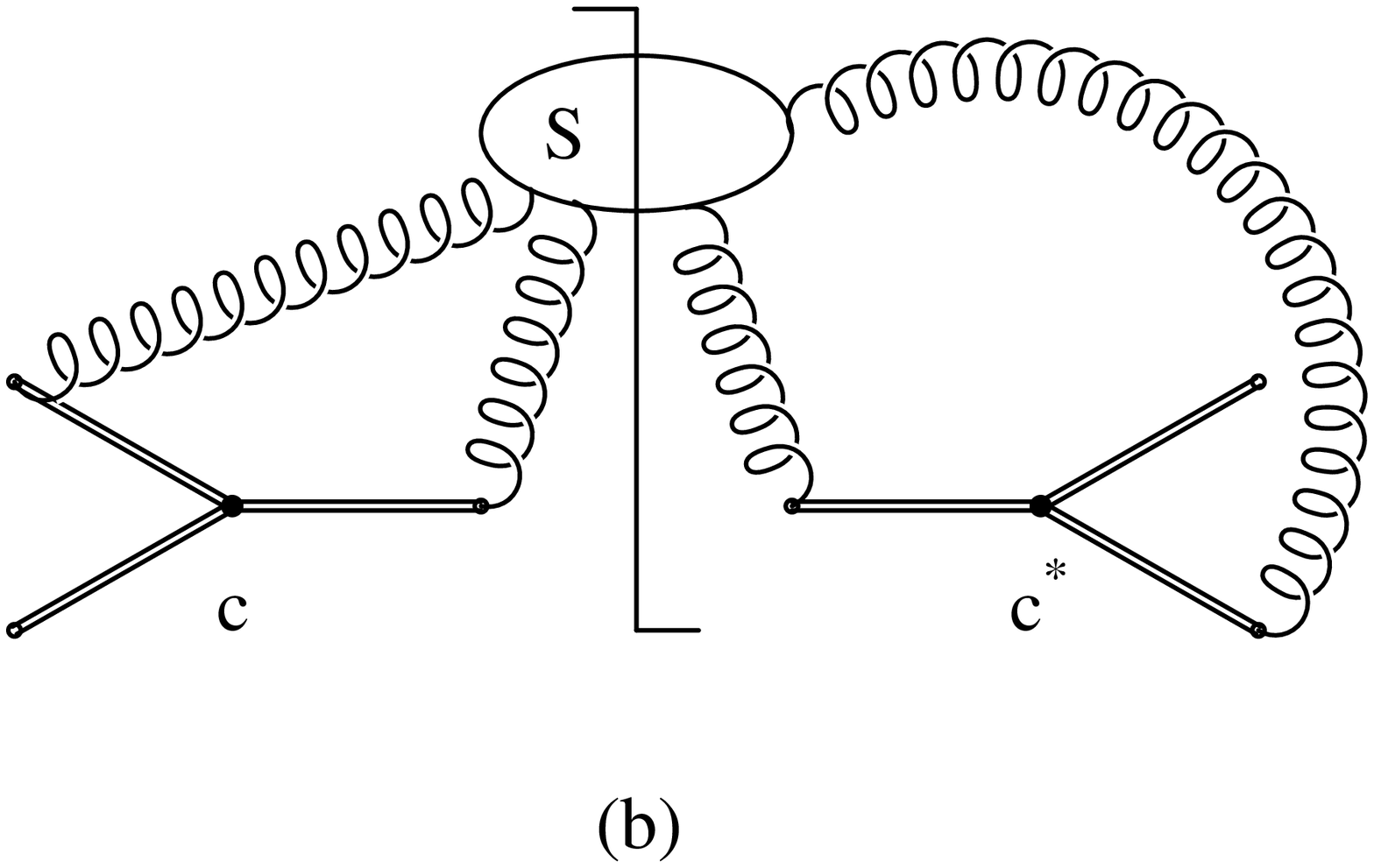,height=1.7in,width=2.5in,clip=}}
{Fig. 1. (a) Factorization for $V$ + jet production 
near partonic threshold.
(b) The soft-gluon function $S$, in which the vertices $c$ link 
eikonal lines representing the partons in the process.}
\label{fig1}
\end{figure}

The resummation of the $N$-dependence of the jet and soft functions
in Eq. (\ref{nresum}) depends on their renormalization properties 
\cite{KS,KOS}. The factor ${\tilde \psi}/{\tilde \phi}$ is universal between 
electroweak and QCD-induced hard processes and its resummation was 
first done in the context of the Drell-Yan process \cite{ster,CT1}. 
It contributes an enhancement to the cross section, while the final-state
jet ${\tilde J}$ gives a negative contribution \cite{KOS,LOS}. 
The UV divergences induced by factorization in the hard and soft 
functions cancel against each other since there are no additional 
UV divergences aside from those already removed through the usual 
renormalization procedure.
The renormalization of the hard and soft functions can be written as
\beq
H^{(0)}= \prod_{i=a,b} Z_i^{-1} Z_S^{-1} H (Z_S^*)^{-1} \, ,
\quad 
S^{(0)}=Z_S^* \, S Z_S \, ,
\label{softren}
\eeq
where $Z_S$ is the renormalization constant of the soft function,
and $Z_i$ is the renormalization constant of the $i$th incoming
partonic field.
Then $S$ satisfies
the renormalization group equation
\beq
\left(\mu\frac{\partial}{\partial\mu}+\beta(g)\frac{\partial}{{\partial}g}
\right)S=-2 ({\rm Re}\, \Gamma_S) \, S\, ,
\quad
\Gamma_S (g)=-\frac{g}{2} \frac {\partial}{{\partial}g}
{\rm Res}_{\epsilon \rightarrow 0} Z_S (g, \epsilon) \, ,
\label{rge}
\eeq
where $\Gamma_S$ is the soft anomalous dimension.
We may then evolve $S$ from the scale $Q/N$ to $\mu_F$ as
\beqa
{\tilde S}\left(\frac{Q}{\mu_F N},\alpha_s(\mu_F^2)\right) 
={\tilde S}\left(1,\alpha_s(Q^2/N^2)\right) \;
\exp \left[\int_{\mu_F}^{Q/N} {d\mu' \over \mu'} \, 
2 {\rm Re} \Gamma_S\left(\alpha_s(\mu'^2)\right)\right] \, .
\nonumber \\
\eeqa
The resummation of the $N$-dependence of each of the functions 
in the refactorized cross section, Eq. (\ref{nresum}), 
in the $\overline{\rm MS}$ factorization scheme, then gives
\cite{KS,KOS,LOS}
\beqa
E_Q\frac{d{\hat\sigma}_{f_af_b\rightarrow V}(N)}{d^3Q} &=&  
\exp \left \{ \sum_{i=a,b} \left [E_{(f_i)}(N_i) \right. \right. 
\nonumber\\ &&  \left. \left. 
-2\int_{\mu_F}^{2p_i \cdot \zeta} {d\mu'\over\mu'}\; 
\left [\gamma_{f_i}(\alpha_s(\mu'{}^2))
-\gamma_{f_if_i}(N_i,\alpha_s(\mu'{}^2)) \right] \right] \right\}
\nonumber \\ && \hspace{-15mm}\times \; 
\exp \left \{E'_{(J)}(N) \right\} \;
H\left(\alpha_s(\mu_F^2)\right) 
\nonumber \\ && \hspace{-15mm} \times \; 
{\tilde S}\left(1,\alpha_s(Q^2/N^2)\right) \;
\exp \left[\int_{\mu_F}^{Q/N} {d\mu' \over \mu'} \, 
2 {\rm Re} \Gamma_S\left(\alpha_s(\mu'^2)\right)\right]\, ,
\label{resxsect}
\eeqa
with $\zeta^{\mu}=p_c^{\mu}/Q$.
The incoming parton-jet anomalous dimensions \cite{KOS,LM}, defined through
\beq
\mu \frac{d{\tilde \psi}_{f/f}(N, Q/\mu,\epsilon)}{d\mu}=
2\gamma_f(\alpha_s(\mu^2)) \, {\tilde \psi}_{f/f}(N, Q/\mu,\epsilon)  
\eeq
and 
\beq
\mu \frac{d{\tilde \phi}_{f/f}(N, \mu^2,\epsilon)}{d\mu}=
2\gamma_{ff}(N,\alpha_s(\mu^2)) \, {\tilde \phi}_{f/f}(N,\mu^2,\epsilon) \, , 
\eeq
are given by
\beqa
\gamma_q &=& {3\over 4} C_F {\alpha_s\over\pi}\,; \qquad
\gamma_{qq} = - \left(\ln{N} - {3\over 4}\right) C_F {\alpha_s\over\pi} \, ,
\nonumber \\ \gamma_g &=& {\beta_0\over 4} {\alpha_s\over\pi}\,; \qquad
\gamma_{gg} = - \left(C_A \ln{N}-{\beta_0\over 4}\right){\alpha_s\over\pi}\, ,
\eeqa
for quark and gluon jets, respectively.
$\beta_0= (11C_A-2n_f)/3$ is the one-loop coefficient of the $\beta$
function, with $n_f$ the number of quark flavors. 
The exponents for the incoming jets are \cite{KS,KOS,LOS}
\beqa
E_{(f_i)}(N_i)
&=&
-\int^1_0 dz \frac{z^{N_i-1}-1}{1-z} \; 
\left \{\int^{1}_{(1-z)^2} \frac{d\lambda}{\lambda} 
A_{(f_i)}\left[\alpha_s\left((2p_i\cdot\zeta)^2\lambda\right)\right] \right.
\nonumber\\ &&  \hspace{15mm} \left.
{}+\frac{1}{2}\nu^{(f_i)}\left[\alpha_s\left((2p_i\cdot\zeta)^2(1-z)^2\right)
\right] 
\right\}\, 
.\label{injet}
\eeqa
At next-to-leading order accuracy in $\ln(N)$, we need 
$\nu^{(f)}=2C_f \alpha_s/\pi$
and $A_{(f)}(\alpha_s) = C_f\left ( {\alpha_s/ \pi} 
+(K/2) \left({\alpha_s/\pi}\right)^2\right)$,
with $C_f=C_F\ (C_A)$ for an incoming quark (gluon), and  
$K= C_A(67/18-\pi^2/6)-5n_f/9$.

The exponent for the final-state jet is \cite{KOS,LOS}
\beqa
E'_{(J)}(N)&=&
\int^1_0 dz \frac{z^{N-1}-1}{1-z} \; 
\left \{\int^{(1-z)}_{(1-z)^2} \frac{d\lambda}{\lambda} 
A_{(J)}\left[\alpha_s(\lambda Q^2)\right] \right.
\nonumber\\ &&  \left. 
{}+ B'_{(J)}\left[\alpha_s\left((1-z)Q^2\right)\right]
+ B''_{(J)}\left[\alpha_s\left((1-z)^2Q^2\right)
\right] \right\}
\label{finjet}
\eeqa
where\footnote{Note that Eq.~(\ref{finjet}) corrects the arguments of
$\alpha_s$ in $B'$ in Eqs. (11) and (12) of Ref.~\cite{LOS}.}
\beqa
B'_{(q)}&=&\frac{\alpha_s}{\pi} \left(-\frac{3}{4}\right) C_F \, ,
\quad
B''_{(q)}=\frac{\alpha_s}{\pi} C_F \left[\ln(2\nu_q)-1\right] \, ,
\nonumber \\ 
B'_{(g)}&=&\frac{\alpha_s}{\pi} \left(- {\beta_0\over 4}\right) \, ,
\quad
B''_{(g)}=\frac{\alpha_s}{\pi} C_A \left[\ln(2\nu_g)-1\right] \, .
\eeqa
The $\nu_i$ terms are gauge dependent. They are defined by
$\nu_i \equiv (\beta_i \cdot n)^2/|n|^2$,
where $\beta_i=p_i\sqrt{2/s}$ are the particle velocities and $n$ 
is the gauge vector, chosen so that $p_i \cdot \zeta=p_i \cdot n$ 
for $i=a,b$ \cite{LOS}.
In our calculations we use Feynman rules for eikonal diagrams
in axial gauge (resummation calculations have also been done in 
covariant gauge \cite{Li}).

The soft anomalous dimensions are calculated explicitly
by evaluating one-loop vertex corrections \cite{KS,NK}.
For the $qg \longrightarrow qV$ channel
in the kinematics (\ref{partkin}) we find \cite{LOS,NK}
\beqa
\Gamma_S^{qg \rightarrow qV}&=&\frac{\alpha_s}{2\pi} 
\left\{C_F\left[2\ln\left(\frac{-u}{s}\right) 
-\ln(4 \nu_{q_a} \nu_{q_c})+2 \right] \right.
\nonumber \\ && \hspace{10mm} \left.
{}+ C_A\left[\ln\left(\frac{t}{u}\right)
-\ln(2\nu_g) +1 -\pi i \right]\right\}\, .
\label{softqg} 
\eeqa
The soft anomalous dimension for $q{\bar q} \longrightarrow gV$ is
\cite{LOS,NK}
\beq
\Gamma_S^{q{\bar q} \rightarrow gV} =\frac{\alpha_s}{2\pi}\left\{C_F\left[
-\ln(4\nu_q \nu_{\bar q})+2 -2 \pi i\right] 
+ C_A\left[\ln\left(\frac{tu}{s^2}\right)
-\ln(2\nu_g) +1 +\pi i \right]\right\}\, .
\label{softqq}
\eeq
Eqs.~(\ref{softqg}) and (\ref{softqq}) coincide with the corresponding
soft anomalous dimensions for direct photon production \cite{LOS,NK}.
Substituting Eqs.~(\ref{injet}) through (\ref{softqq}) in the resummed
cross section, Eq.~(\ref{resxsect}), we see that at NLL accuracy all 
the gauge-dependent terms cancel out in the exponent, for both the
$qg \longrightarrow qV$ channel and the
$q{\bar q} \longrightarrow gV$ channel.

We can rewrite the resummed cross section 
in a form which is more convenient
for the calculation of the fixed-order expansions.
Using the renormalization group behavior of the product $HS$
from Eq. (\ref{softren}),
\beq
\mu \frac{d}{d\mu} \ln \left[H(\mu) S(Q/(N \mu))\right]=
-2\left[\gamma_a(\alpha_s(\mu^2)+\gamma_b(\alpha_s(\mu^2))\right] \, ,
\eeq
and the relation
\beq
H(\alpha_s(Q^2))=H(\alpha_s(\mu_R^2))
\exp\left[\int_{\mu_R}^Q 2 \frac{d\mu'}{\mu'}
\beta(\alpha_s(\mu^2))\right] \, ,
\eeq
with $\beta(\alpha_s) \equiv \mu d\ln g/d\mu = - \beta_0 \alpha_s/4\pi
+ ...$, we can write Eq.~(\ref{resxsect}) as
\beqa
E_Q\frac{d{\hat\sigma_{f_af_b\rightarrow V}}(N)}{d^3Q} &=&
H\left(\alpha_s(\mu_R^2)\right) 
\exp\left[2\int_{\mu_R}^Q \frac{d\mu'}{\mu'} \beta(\alpha_s(\mu'^2))\right]  
\nonumber \\ && \hspace{-35mm} \times \;
\exp \left \{ \sum_{i=a,b} \left [E_{(f_i)}(N_i)  
-2\int_{\mu_F}^{2p_i \cdot \zeta}{d\mu'\over\mu'}\; 
\left [\gamma_{f_i}(\alpha_s(\mu'{}^2))
-\gamma_{f_if_i}(N_i,\alpha_s(\mu'{}^2)) \right] \right] \right\}
\nonumber \\ &&  \hspace{-28mm}  \times \; 
\exp \left \{E'_{(J)}(N) \right\} \;
\exp \left[2\int_{\mu_F}^Q {d\mu' \over \mu'} 
\left(\gamma_a(\alpha_s(\mu'^2))
+\gamma_b(\alpha_s(\mu'^2))\right)\right]
\nonumber \\ && \hspace{-28mm} \times \;   
{\tilde S}\left(1,\alpha_s(Q^2/N^2)\right) \;
\exp \left[\int_{Q}^{Q/N} {d\mu' \over \mu'} \, 
2 {\rm Re} \Gamma_S\left(\alpha_s(\mu'^2)\right)\right] \, .
\label{resum}
\eeqa
                
\mysection{Next-to-next-to-leading order expansion of the resummed
cross section}

In this section, we expand the NLL resummed cross section up to NNLO, 
invert back from moment space to momentum space,
and perform a comparison with the NLO results of 
Ref.~\cite{AR,gpw}. At the parton level, the subprocesses for the 
production of an electroweak boson and a parton are given in 
Eq.~(\ref{partsub}).

We can write the NLO corrections from Eq. (\ref{resum}) for 
$qg \longrightarrow qV$ 
in single-particle inclusive kinematics at NLL accuracy as
\beqa
E_Q\frac{d{\hat\sigma}^{\rm NLO}_{qg \rightarrow qV}}{d^3Q} &=& 
\sigma^B_{qg \rightarrow qV} 
{\alpha_s(\mu_R^2)\over\pi}\,
\left\{ (C_F+2C_A) \left[\frac{\ln(s_2/Q^2)}{s_2}\right]_+ \right.  
\nonumber \\ && \hspace{-25mm}
{}-\left[\left(C_F + C_A\right) \ln\left(\frac{\mu_F^2}{Q^2}\right)
+ {3\over 4} C_F +  C_A \ln{\left(t u \over s Q^2\right)}\right] 
\left[\frac{1}{s_2}\right]_+
\nonumber \\ && \hspace{-25mm}  
{}+\delta(s_2) \left[\ln\left(\mu_F^2\over Q^2\right)
\left(-\frac{\beta_0}{4}+C_F\left(\ln\left(\frac{-u}{Q^2}\right)
-\frac{3}{4}\right)+C_A\ln\left(\frac{-t}{Q^2}\right)\right) \right.
\nonumber \\ && \hspace{-20mm}  \left. \left.
{}+\frac{\beta_0}{4} \ln\left(\mu_R^2\over Q^2\right)\right]\right\} \, .
\nonumber \\
\label{qgnlo}
\eeqa
Here the Born term is
\beqa
\sigma^B_{qg \rightarrow qV} &=& \frac{\alpha \alpha_s(\mu_R^2)C_F}{s(N_c^2-1)}
A^{qg} \, \sum_{f} (|L_{ff_a}|^2+|R_{ff_a}|^2) \, ,\\
A^{qg} &=& - \left(\frac{s}{t}+\frac{t}{s}+\frac{2uQ^2}{st}\right) \, , 
\nonumber
\eeqa
with $L$ and $R$ the left- and right-handed couplings of the
electroweak boson to the quark line, $f$ the quark flavor and $\sum_f$
the sum over the flavors allowed by the CKM mixing and by the energy
threshold. We choose for the $L\,,R$ couplings the conventions of
Ref.~\cite{gpw}. The Born differential cross section is
$E_Q d\sigma^B_{qg\rightarrow qV}/d^3Q
=\sigma^B_{qg \rightarrow qV} \, \delta(s_2)$.

We have kept the factorization and renormalization scales separate.
The  expansion gives all $[\ln(s_2/Q^2)/s_2]_+$
and $[1/s_2]_+$ terms but only scale-dependent $\delta(s_2)$ terms.

Our one-loop expansion can be compared with the exact NLO cross 
section~\cite{gpw}
in the proximity of the partonic threshold, $s_2\to 0$, i.e. with the
singular terms of the type (\ref{zplus}) in the contribution of 
the subprocess $qg \longrightarrow qVg$ to the cross section for 
$qg \longrightarrow qV$. We find full agreement\footnote{Note that 
Gonsalves et al.~\cite{gpw} use
``A+'' distributions, which are related to our ``+'' distributions by
$[\ln^n(s_2/Q^2)/s_2]_+=[\ln^n(s_2/Q^2)/s_2]_{A^+} +(1/(n+1))\ln^{n+1}(A/Q^2) 
\delta(s_2)$.} with the results of 
Ref.~\cite{gpw}. 

Then we expand the resummed cross section, Eq.~(\ref{resum}), for 
$qg \longrightarrow qV$ to two-loop order and compute
the NNLO corrections
in single-particle inclusive kinematics at NLL accuracy,
including all the factorization and renormalization scale 
dependent terms. We obtain
\beqa
E_Q\frac{d{\hat\sigma}^{\rm NNLO}_{qg \rightarrow qV}}{d^3Q}&=& 
\sigma^B_{qg \rightarrow qV}
\left({\alpha_s(\mu_R^2)\over\pi}\right)^2 \left\{ 
\frac{1}{2}\left(C_F+2C_A\right)^2 \left[\frac{\ln^3(s_2/Q^2)}{s_2}\right]_+
\right.
\nonumber \\ && \hspace{-32mm}
{}+\left[-\frac{3}{2} (C_F+2C_A) 
\left(\left(C_F + C_A\right) \ln\left(\mu_F^2\over Q^2\right) 
+ {3\over 4} C_F + C_A \ln\left(t u \over s Q^2 \right) \right) \right.  
\nonumber \\ &&  \hspace{10mm}\left.
{}- {\beta_0\over 2} \left({C_F \over 4}+C_A\right)\right] 
\left[\frac{\ln^2(s_2/Q^2)}{s_2}\right]_+ 
\nonumber \\ && \hspace{-32mm}
{}+\ln\left(\frac{\mu_F^2}{Q^2}\right)
\left[\left(C_F + C_A\right)^2\ln\left(\frac{\mu_F^2}{Q^2}\right) 
+ \frac{3}{4}C_F^2+ 2C_A(C_F+C_A) \ln\left(\frac{tu}{sQ^2}\right) \right. 
\nonumber \\ &&  \hspace{-25mm}\left.
{}+(C_F+2C_A) \left(C_F\ln\left(\frac{-u}{Q^2}\right) 
+C_A\ln\left(\frac{-t}{Q^2}\right)-\frac{\beta_0}{4}\right)\right]
\left[\frac{\ln(s_2/Q^2)}{s_2}\right]_+ 
\nonumber \\ && \hspace{-32mm}
{}+\ln\left(\frac{\mu_R^2}{Q^2}\right) \frac{\beta_0}{2}(C_F+2C_A)
\left[\frac{\ln(s_2/Q^2)}{s_2}\right]_+ 
\nonumber \\ &&  \hspace{-32mm}
{}+\ln^2{\left(\mu_F^2\over Q^2\right)}\, \left(C_F + C_A\right)  
\left[C_F \left(-\ln\left({-u \over Q^2}\right) +{3\over 4}\right) 
- C_A \ln\left({-t\over Q^2}\right)  
+{3\beta_0\over 8} \right] \left[\frac{1}{s_2}\right]_+ 
\nonumber \\ && \hspace{-25mm} \left.
{}-\frac{\beta_0}{2} \, (C_F+C_A) \,
\ln{\left(\mu_R^2\over Q^2\right)} \, \ln{\left(\mu_F^2\over Q^2\right)} 
\, \left[\frac{1}{s_2}\right]_+ \right\} \, .
\label{qgnnlo}
\eeqa
Our expansion gives all $[\ln^3(s_2/Q^2)/s_2]_+$
and $[\ln^2(s_2/Q^2)/s_2]_+$ terms but only scale-dependent terms
for $[\ln(s_2/Q^2)/s_2]_+$  and $[1/s_2]_+$. However we can derive 
all the $[\ln(s_2/Q^2)/s_2]_+$ terms by matching with the exact 
NLO cross section\footnote{In Eq.~(2.12) of Ref.~\cite{gpw} there is a typo:
the term $C_2^{qg}$ should be replaced by 
$\delta(s_2)C_2^{qg}n_f + C_3^{qg}$.}. 
Including the $\mu_{R,F}$-dependent
terms given above, the full $[\ln(s_2/Q^2)/s_2]_+$ terms are 
\beqa
&& \sigma^B_{qg \rightarrow qV}
\left({\alpha_s(\mu_R^2)\over\pi}\right)^2
\left\{\frac{1}{2A^{qg}}(C_F+2C_A)\left[B_1^{qg}+B_2^{qg} n_f
+C_1^{qg}+C_2^{qg} n_f \right. \right.
\nonumber \\ && \hspace{20mm} \left. 
+B_3^{qg} {(L_{f_af_a}-R_{f_af_a})\sum_{f}(L_{ff}-R_{ff}) \over
\sum_{f} (|L_{ff_a}|^2+|R_{ff_a}|^2)} \right]
\nonumber \\ && 
{}+\left[(C_F+C_A)\ln\left(\frac{\mu_F^2}{Q^2}\right)
+\frac{3}{4}C_F
+C_A \ln \left(\frac{tu}{sQ^2}\right)\right]^2 
\nonumber \\ && 
{}+(C_F+2C_A) \frac{\beta_0}{4} \ln\left(\frac{\mu_R^2}{Q^2}\right)
+\frac{1}{2} K (C_F+2C_A)-\zeta_2(C_F+2C_A)^2
\nonumber \\ &&  \left.
{}+\beta_0 {C_F\over 4} \left(\frac{3}{4} - \ln\left({s\over Q^2}\right)
\right)
+\beta_0 {C_A\over 2} \ln\left(\frac{tu}{sQ^2}\right) 
\right\} 
\left[\frac{\ln(s_2/Q^2)}{s_2}\right]_{+}\, ,
\eeqa
with $\zeta_2=\pi^2/6$, and with $B_1^{qg}$, $B_2^{qg}$, 
$B_3^{qg}$, $C_1^{qg}$, and $C_2^{qg}$ as given in the Appendix
of Ref. \cite{gpw} but without the renormalization counterterms
and using $f_A \equiv\ln(A/Q^2)=0$. 

Next, we consider the $q{\bar q} \longrightarrow gV$
partonic subprocess, and compute 
the NLO corrections from Eq. (\ref{resum}) 
in single-particle inclusive kinematics at NLL accuracy,
\beqa
E_Q\frac{d{\hat\sigma}^{\rm NLO}_{q{\bar q} \rightarrow gV}}{d^3Q} &=&
{\sigma^B_{q{\bar q} \rightarrow gV}} {\alpha_s(\mu_R^2)\over\pi}\,
\left\{(4C_F-C_A) \left[\frac{\ln(s_2/Q^2)}{s_2}\right]_+  \right.
\nonumber \\  && \hspace{-20mm}
{}- \left[ 2 C_F \ln\left({\mu_F^2\over Q^2}\right) 
+ \left(2 C_F- C_A \right) \ln\left({t u \over s Q^2}\right) 
+\frac{\beta_0}{4}\right] \left[\frac{1}{s_2}\right]_+ 
\nonumber \\  && \hspace{-20mm} \left.
{}+\delta(s_2) \left[\ln\left({\mu_F^2\over Q^2}\right)  
C_F\left(\ln\left(\frac{tu}{Q^4}\right)-\frac{3}{2}\right)
+\frac{\beta_0}{4} \ln\left({\mu_R^2\over Q^2}\right)\right]
\right\}.
\label{qqbarnlo}
\eeqa
Here the Born term is
\beqa
\sigma^B_{q{\bar q} \rightarrow gV} &=&\frac{\alpha \alpha_s(\mu_R^2)C_F}{sN_c}
A^{q\bar q}\,  (|L_{f_bf_a}|^2+|R_{f_bf_a}|^2) \, , \\
A^{q\bar q} &=& \frac{u}{t}+\frac{t}{u}+\frac{2Q^2s}{tu} \, ,
\nonumber
\eeqa
and the Born differential cross section is
$E_Q d\sigma^B_{q {\bar q}\rightarrow g V}/d^3Q
=\sigma^B_{q{\bar q} \rightarrow gV} \, \delta(s_2)$.

The one-loop term can be compared with the
singular terms of the type (\ref{zplus}) in the contribution of 
the subprocesses $q{\bar q} \longrightarrow gVg$  and
$q{\bar q} \longrightarrow Vq{\bar q}$ to the cross section for 
$q{\bar q} \longrightarrow gV$. We are in agreement with Ref.~\cite{gpw}.

The NNLO corrections from Eq. (\ref{resum})
for $q{\bar q} \longrightarrow gV$ 
in single-particle inclusive kinematics at NLL accuracy,
including all the factorization and renormalization scale 
dependent terms, are
\beqa
E_Q\frac{d{\hat\sigma}^{\rm NNLO}_{q{\bar q} \rightarrow gV}}{d^3Q}&=&
{\sigma^B_{q{\bar q} \rightarrow gV}}
\left({\alpha_s(\mu_R^2)\over\pi}\right)^2\,
\left\{ \frac{1}{2}(4C_F-C_A)^2\, 
\left[\frac{\ln^3(s_2/Q^2)}{s_2}\right]_+ \right.
\nonumber \\ &&  \hspace{-20mm}
{}+\left[-\frac{3}{2} (4C_F-C_A)\left(2 C_F \ln\left({\mu_F^2\over Q^2}\right) 
+ \left( 2 C_F - C_A \right) \ln\left({t u \over s Q^2}\right) \right) \right.
\nonumber \\ && \left.
{}-\frac{\beta_0}{2} \left(5C_F -{3\over 2} C_A \right) \right]
\left[\frac{\ln^2(s_2/Q^2)}{s_2}\right]_+   
\nonumber \\ && \hspace{-20mm}
{}+\ln\left({\mu_F^2\over Q^2}\right)
\left[ C_F \left(4C_F \ln\left({\mu_F^2\over Q^2}\right) 
+ \left(4C_F-C_A\right)
\left( \ln\left({tu\over Q^4}\right) - {3\over 2}\right) \right. \right.
\nonumber \\ &&  \left. \left.
{}+ 4\left( 2 C_F- C_A \right) \ln\left({tu \over s Q^2}\right)
\right)+C_F \beta_0 \right] 
\left[\frac{\ln(s_2/Q^2)}{s_2}\right]_+ 
\nonumber \\ && \hspace{-20mm} 
{}+\ln\left({\mu_R^2\over Q^2}\right) \frac{\beta_0}{2}(4C_F-C_A)
\left[\frac{\ln(s_2/Q^2)}{s_2}\right]_+
\nonumber \\ && \hspace{-20mm}
{}+\ln^2\left({\mu_F^2 \over Q^2}\right) \,
C_F \left[C_F\left(-2\ln\left({t u\over Q^4}\right) + 3\right)
+\frac{\beta_0}{4}\right] \left[\frac{1}{s_2}\right]_+ 
\nonumber \\ && \hspace{-20mm} \left.
{}-\beta_0 \, C_F \, 
\ln{\left(\mu_R^2\over Q^2\right)} \, \ln{\left(\mu_F^2\over Q^2\right)} 
\, \left[\frac{1}{s_2}\right]_+\right\}\, .
\label{qqbarnnlo}
\eeqa
Again, only scale-dependent terms
for $[\ln(s_2/Q^2)/s_2]_+$  and $[1/s_2]_+$ are included in 
Eq.~(\ref{qqbarnnlo}).
The full $[\ln(s_2/Q^2)/s_2]_+$ terms are found by matching 
with the exact NLO cross section, 
\beqa
&& \hspace{-5mm}{\sigma^B_{q{\bar q} \rightarrow gV}}
\left({\alpha_s(\mu_R^2)\over\pi}\right)^2
\left\{\frac{1}{2A^{q \bar q}}(4C_F-C_A)\left[B_1^{q \bar q}+C_1^{q \bar q}
+(B_2^{q \bar q}+D_{aa}^{(0)}) \, n_f \right. \right.
\nonumber \\ && \hspace{15mm} \left.
+B_3^{q \bar q} \delta_{f_af_b} {(L_{f_af_a}-R_{f_af_a})
\sum_{f}(L_{ff}-R_{ff}) \over (|L_{f_bf_a}|^2+|R_{f_bf_a}|^2)} \right] 
\nonumber \\ && 
{}+\left[2C_F\ln\left(\frac{\mu_F^2}{Q^2}\right)
+(2C_F-C_A)\ln\left(\frac{tu}{sQ^2}\right)+\frac{\beta_0}{4}\right]^2
\nonumber \\ && 
{}+(4C_F-C_A)\frac{\beta_0}{4}\ln\left(\frac{\mu_R^2}{Q^2}\right)
+(4C_F-C_A)\frac{K}{2}+\frac{\beta_0^2}{16}-\zeta_2(4C_F-C_A)^2
\nonumber \\ &&  \left.
{} +\beta_0
\left(C_F- {C_A\over 2}\right) \ln\left(\frac{tu}{sQ^2}\right)
- C_A\, {\beta_0\over 4} \ln\left(s\over Q^2\right)
\right\} \, \left[\frac{\ln(s_2/Q^2)}{s_2}\right]_{+} \, ,
\eeqa
with $B_1^{q \bar q}$, $B_2^{q \bar q}$, $B_3^{q \bar q}$,
$C_1^{q \bar q}$, and $D_{aa}^{(0)}$
as given in the Appendix of Ref. \cite{gpw} but without the renormalization 
counterterms and using $f_A=0$. 

\mysection{Conclusion}
We have presented the resummed cross section at NLL accuracy
for electroweak-boson production at large transverse momentum
near partonic threshold. 
The expansion of the resummed cross section at NLO agrees with previous
exact NLO results while the NNLO expansion provides new predictions
at higher order. In future work we plan to study the numerical 
significance of resummation for $W$+ jet production. Related studies
for other processes have shown reduced factorization scale-dependence,
expected on theoretical grounds \cite {EPIC99},
and increases over the NLO cross section.

\mysection*{Acknowledgements}

We thank Stefano Catani, Eric Laenen, Jeff Owens,
and George Sterman for useful discussions.
The work of N.K. was supported in part by the U.S. Department of Energy.

\end{document}